\documentclass[aps, pra,twocolumn,showpacs,amsmath,amssymb,superscriptaddress, longbibliography, 10pt]{revtex4-1}

\usepackage{tgtermes}
\usepackage[utf8]{inputenc}
\usepackage{bm}
\usepackage{color}
\usepackage{makeidx}
\usepackage{amsfonts}
\usepackage{amssymb}
\usepackage{amsthm}
\usepackage{graphicx}
\usepackage{epsfig}
\usepackage{subfigure}
\usepackage{amsmath,amssymb}
\usepackage[colorlinks=true,citecolor=blue,urlcolor=blue]{hyperref}
\usepackage{color}

\newcommand{\bea}{\begin{eqnarray}}
\newcommand{\ea}{\end{eqnarray}}
\newcommand{\eea}{\end{eqnarray}}

\newcommand{\sumint}[1]

\begin{document}

\title{Controllable splitting dynamics of a doubly quantized vortex in a rotating ring-shaped condensate}

\author{Bo Xiong}
\affiliation{School of Science, Wuhan University of Technology, Wuhan 430070, China}
\author{Tao Yang}
\email{yangt@nwu.edu.cn}
\affiliation{Institute of Modern Physics, Northwest University, Xi'an, 710069, China}
\affiliation{Shaaxi Key Laboratory for Theoretical Physics Frontiers, Xi'an, 710069, China}
\author{Yu-Ju Lin}
\affiliation{Institute of Atomic and Molecular Sciences, Academia Sinica, Taipei 10617, Taiwan, Republic of China}
\author{Daw-wei Wang}
\affiliation{Institute of Physics, National Center for Theoretical Science, Hsingchu, Taiwan 300, Republic of China}
\affiliation{Department of Physics, National Tsing-Hua University, Hsinchu, Taiwan 300, Republic of China}

\date{\today}

\begin{abstract}
    We study the dynamics of a doubly quantized vortex (DQV), created by releasing a ring-shaped Bose-Einstein condensate 
with quantized circulation into harmonic potential traps.
It is shown that 
a DQV can be generated and exists stably in the middle of the ring-shaped condensate with the initial 
circulation $s = 2$ after released into the rotationally symmetric trap potential. 
For an asymmetric trap with a small 
degree of anisotropy the DQV initially splits into two singly quantized vortices and revives again
but eventually evolves into two unit vortices due to the dynamic instability. For the degree of anisotropy 
above a critical value, the DQV is extremely unstably and decays rapidly into two singlet vortices. 
The geometry-dependent lifetime of the DQV and vortex-induced excitations are also discussed intensively.
\end{abstract}


\maketitle
\section{Introduction}

	 The realization of Bose-Einstein condensates (BECs) provides a controllable platform to investigate 
the properties of superfluid. One of the hallmarks of superfluid is the quantization of fluid circulation, 
generally manifested through the formation of quantum vortices. Large arrays of singly quantized vortices with a 
$2\pi$ phase singularity around the core region have been observed in rapidly rotating
clouds of cold alkali metal atoms confined in a magnetic trap of cylindrical symmetry \cite{Madison:2000a, Shaeer:2001a,
Hodby:2001a}. The condensate can also acquire high angular momentum which makes a giant vortex with a $2s \pi$
phase singularity appear, where the integer $s \geq 2$ is the quantum number called the charge or winding number of 
the vortex. At large rotational speeds, the vortex lattices in a rotating hard-walled bucket are not stable and 
collapse to form a giant vortex state \cite{Fischer:2003a}. The giant vortices with quantum number larger 
than unity have been created in atomic gases by flipping of a magnetic field 
\cite{Leanhardt:2002a, Shin:2004a, Kumakura:2006a, Isoshima:2007a, Mottonen:2007a} and   by using a Laguerre-Gaussian laser beam 
\cite{Andersen:2006a}. 

Generally, giant vortices are energetically unstable in spinless, macroscopic, and homogeneous
superfluids, because the energy of a vortex depends on the square of the winding number $s$, and under the condition 
of total angular momentum conservation the system prefers to $s$ singly quantized vortices rather than one $s$-quantized vortex. 
Multiply quantized vortices are also dynamically unstable by transferring the kinetic energy to coherent 
excitation modes which is driven by interatomic interactions. Therefore, it will be a challenge and interesting 
issue to create a stable multiply quantized vortex with relatively long lifetime and investigate 
some complex dynamics associated with such quantum circulation in a controllable way. It is shown that a stable giant vortex may 
be realized in condensates with a localized pinning potential \cite{Simula:2002a}, 
in a quartic potential \cite{Lundh:2002a}, or oblate condensates \cite{Rabina:2018a}. In harmonically trapped 
three-dimensional (3D) condensates the stability of a giant vortex depends strongly on the trap anisotropy and the 
strength of interatomic interaction. The splitting instability of multiply quantized vortices can be suppressed in
some narrow parameter range \cite{Rabina:2018a, Huhtamaki:2006a, Lundh:2006a}.

   Ring-shaped BECs are, as a multiply connected structure, more stable for the superfluid behavior than singly connected BECs. 
They are widely employed to investigate a variety of issues, including persist current \cite{Ryu:2007a, Beattie:2013a, Yakimenko:2013a},
weak connection \cite{Ramanathan:2011a, Wright:2013a, Piazza:2009a, Piazza:2013a}, generation of quasi-stable dark-soliton structure
\cite{Gallucci:2016a}, and mimicking cosmic expansion \cite{Eckel:2018a}. 
Recent experiments have achieved to realize multiply quantized circulation in a ring-shaped BEC \cite{Ryu:2007a, Moulder:2012a, Murray:2013a}. This implies that 
it is applicable to produce multiply quantized vortex in a rotating toroidal BEC by releasing it into some certain geometric traps. 
The multiply quantized vortex would be stabilized by the toroidal BEC because it costs too much energy for the vortex core to move from
the center of the torus, where the density is zero, through the high density atomic cloud \cite{Leggett:2001a, 
Bloch:1973a, Tempere:2001a}. 
On the other hand, due to the inward motion of the ring-shaped BEC, the cloud around the hole may bound the vortex core and suppress 
the dynamical instability. 
Thus, it is highly promising to explore the dynamics of a multiply quantized vortex in the ring-shaped BEC.
The dynamics of such giant vortices is also an interesting topic, and multiquantum vortices could be
used to implement a ballistic quantum switch \cite{Melnikov:2002a} or realize bosonic quantum Hall states 
\cite{Roncaglia:2011a}. Moreover, the vortex splitting can be employed for generating quantum turbulence with 
controllable net circulation \cite{Abraham:1995a, Aranson:1996a, Cidrim:2016a}.


   In this paper, we study the dynamics of a doubly quantized vortex (DQV) in a ring-shaped BEC released in a harmonic trap
varying from rotationally symmetric geometry to asymmetric one. The stability of the DQV and its lifetime with respect
to the trap asymmetry and the interatomic interaction are explored. The associated characteristic dynamics 
are also shown. Moreover, the higher-fold vortex structures excited by the splitting process of the initial DQV with
increasing interatomic interaction of the condensate are also discussed. 

	
	%
	%
	%
	%
	%
	%
	%

	
\section{Simulation schemes and numerical methods}

   Our general protocol involves preparing a ring-shaped clouds with doubly quantized circulation, of which
wave function can be described in the cylindrical coordinate system by
   \begin{equation} \label{wave1}
			\Psi (r, \phi, z) = \left\{ \begin{array}{cc} A \sqrt{ \mu - \frac{1}{2} m \omega_{r_0}^{2} r^{2}} f(r, z)  e^{i 2\phi}  &  \mu \geq \frac{1}{2} m \omega_{r_0}^{2} r^{2} \\
                                  0     &  {\rm otherwise}
                                     \end{array} \right.
   \end{equation}
where $A$ is the normalization coefficient. The distribution function 
   \begin{equation} \label{wave2}
			f (r, z)= \left[1 - e^{- a \frac{r^{2}}{2d^{2}} e^{- \frac{r^{2}}{2d^{2}}}} \right] e^{- \frac{z^2}{2 \sigma_{z}^{2}}},
	 \end{equation}
with $\sigma_z = \sqrt{\hbar/m \omega_z}$ \footnote{We also use the trial wave function, $\psi(r, \phi, z) = A \sqrt{\mu - \frac{1}{2} m
\omega_{r_0}^{2} (r - r_{M})^2}\exp[- z^2/2\sigma_{z}^{2}]$ for $\mu \geq \frac{1}{2} m \omega_{r_0}^{2} (r - r_{M})^2$ and otherwise 0,
to match the ring-shaped BEC, which is created by a order-1 Laguerre-Gaussian beam with the form $V_{0} (r, z) = - K I_{0}
\left( \frac{r}{r_{0}} \right)^{2} \exp\left[ - \frac{2 r^{2}} {r_{0}^{2}} \right] + \frac{1}{2} \omega_{z}^{2} z^{2}$. For a
typical laser field, $I_{0} = 4 P_{\rm LG}/\pi r_{0}^{2}$ and $K = h \cdot 3072 \rm{k Hz} \frac{(\mu m)^{2}}
{m W}$. $r_0$ is the beam waist and can be focused to as small as a few $\mu$m under diffraction limit.
$r_{M}$ is the minimum point of trap potential along the radius direction, where the first part of the order-1 Laguerre-Gaussian beam
can be treated approximately as $\frac{1}{2} \omega_{r_0}^{2} (r - r_{M})^{2}$ with
$\omega_{r_0} = \sqrt{ \frac{16 K P_{\rm LG}} {\pi e m r_{0}^{4}}}$. We emphasize that no qualitative difference in the dynamics is found
in our simulation for two types of initial wave functions.}. This wave function is employed in some experiments to map the ring-shaped cloud.
Such a BEC is created by the toroidal trap potential, which can be produced by using a blue-detuned laser beam to make a repulsive
potential barrier in the middle of a harmonic magnetic trap \cite{Ryu:2007a}. The toroidal trap potential has the general form,
\begin{equation} \label{trap1}
   V_0(r, z) = \frac{1}{2}m \omega_{r_0}^{2} r^2 + \frac{1}{2} m \omega_{z}^{2} z^2  + V_{0}
	 \exp\left[ - \frac{2 r^2}{d_{0}^{2}} \right]
\end{equation}
where $\omega_{r_0}$ and $\omega_{z}$ are the harmonic trapping frequencies along $r$ and $z$ direction, respectively.
$V_0$ is the maximum optical potential, proportional to the laser intensity. A BEC density distribution with a hole
in the center is created by the plug beam, if the chemical potential, $\mu$, is less than $V_0$. The parameter $a$ in 
the distribution function (\ref{wave2}) is tunable according to the intensity of the laser beam to determine the size of 
the hole. We choose such experimental setup because the toroidal trap can be easily changed into a harmonic traps by deceasing 
the laser power. The harmonic trap frequencies can also be changed by adjusting the magnetic field.

   The initial wave function employed in our simulation originates from an experimental data \footnote{Private communication} 
which refers to a ring-shaped condensate with $12\mu$m radial length and $6.5\mu$m hole size. To realize such a $^{87}$Rb 
ring-shaped BEC with the total atom number $N = 10^5$, the radial and axial frequencies are chosen
to be $\omega_{r_0} = 2\pi \times 47$ Hz and $\omega_z = 2\pi \times 238$ Hz. A typical density profile of the ring-shaped condensate is
given in Fig.\,\ref{Fig1}(a).

In the subsequent process, the prepared cloud is released into a harmonic trap with different symmetry, varying 
from isotropic geometry in the radial direction to anisotropic geometry. Then we examine the role of the geometry of the trap
on the dynamics of the system, which is governed by the Gross-Pitaevskii (GP) equation
   \begin{equation} \label{GPE1}
	    i\hbar \frac{\partial \Psi}{\partial t} = - \frac{\hbar^{2}}{2m} \nabla^{2}\Psi + V \Psi + g|\Psi|^{2} \Psi,
	 \end{equation}
where $V = \frac{1}{2} \omega_{r}^{2} r^{2} \left[1 - \eta \sin^{2}\phi \right] + \frac{1}{2}\omega_{z}^{2} z^{2}$ with
$\omega_{r} = 2\pi \times 79$Hz. The parameter $\eta$ indicates the degree of anisotropy
in the trap geometry. We set $0 \leq \eta \leq 1$ where $\eta = 0$ accounts for the isotropic trap and $\eta = 1$ for the quadratic order
potential along $x$ direction while uniform along $y$ direction.  Here the interatomic coupling coefficient $g$ can be tuned from
$0.1g_0$  to $50g_0$ where $g_0 = 4\pi \hbar^{2} a_{s}/m$. $a_s$ is set to be $5.4$nm and $m = 1.44 \times 10^{-25}$kg, appropriate to a $^{87}$Rb
condensate.

	  To explore time evolution of the doubly quantized vortex created by a ring-shaped BEC, the traditionally prevalent way is to numerically 
solve the 3D GP equation (\ref{GPE1}) in the Cartesian coordinate system. Note that at the merging point around $r = 0$, the momentum of the 
ring-shaped BEC is relatively high and also the DQV formed in the merging area is strongly sensitive to the asymmetric-geometry 
potential. This indicates that a sufficiently small grid size is desired to avoid the distortion arisen from the square lattice 
in the Cartesian coordinate system. Thus a very demanding computational task is required. Taking the ring-shaped geometry into 
account and to save the calculation time, we solve the GP equation by using the split-step Crank-Nicolson algorithm on the 
3D cylindrical coordinate system with $r, \phi, z$ spatial grid of $320 \times 200 \times 100$ points. Then we transform the results 
in the cylindrical coordinate system into ones in the Cartesian coordinate system for more intuitive understanding. We stress that 
to achieve the stable results for the dynamics of DQV produced in the ring-shaped BEC, our calculation in the $r, \phi, z$ space is 
much less time-consuming than the one in the $x, y, z$ space, e.g., at least five times less than the calculation time in the 
Cartesian coordinate system for one dynamics process. To decrease the error induced by the square lattice in the dynamics of 
giant vortices,  the latest work employs discrete exterior calculus with tetrahedral tiling \cite{Rabina:2018a}. 

\section{Stable doubly quantized vortex}

\begin{figure}[hptb]
\centering
\includegraphics[scale = 0.12]{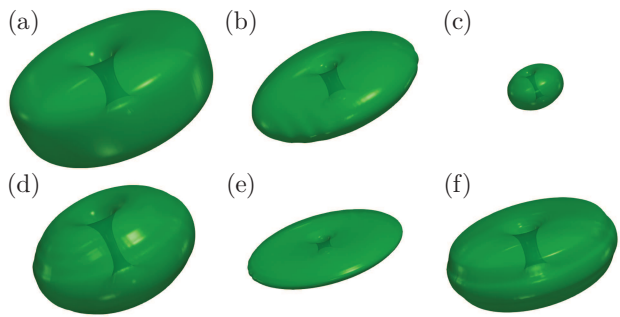} 		        	
\caption{(Color online) Temporal density isosurface of a ring-shaped condensate with the circulation $s = 2$ in a symmetric trap,
i.e., $\eta = 0$ for $1\%$ of the maximum density and $g = g_0$. Different panels correspond to different time: $t = 0$ms (a), 0.57ms (b), 1.1ms (c), 1.7ms (d), 2.0 ms (e), 2.6 ms (f). }
\label{Fig1}
\end{figure}

\begin{figure*}[t]
\centering
\includegraphics[scale = 0.15]{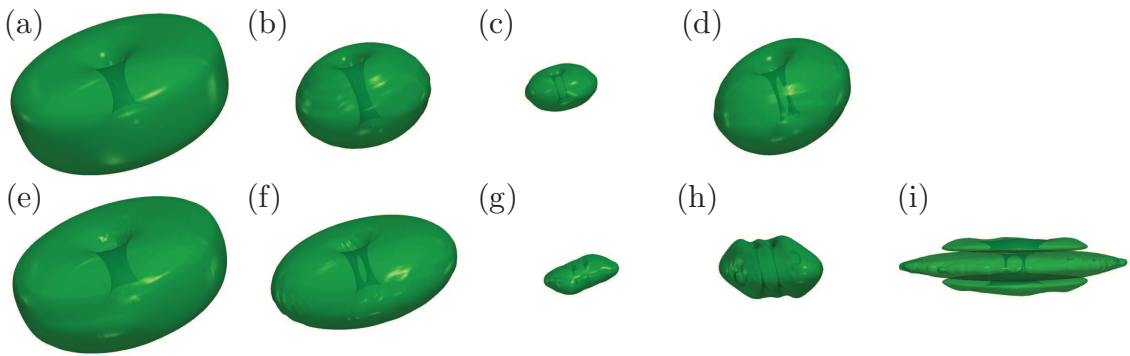} 		        	
\caption{(Color online) Temporal density isosurface of the ring-shaped condensate in an asymmetric trap, i.e., $\eta = 0.1$ for
$1\%$ of the maximum density and $g = g_0$. Different panels correspond to different time: $t = 0$ms (a),
0.85ms (b), 1.1ms (c), 1.7ms (d), 2.6ms (e), 3.1ms (f), 3.7ms (g), 6ms (h), 7ms (i).}
\label{Fig2}
\end{figure*}

To get first insight, and see the stability of a DQV, we begin by discussing the oscillation of the ring-shaped BEC
with the symmetric density envelope in an isotropic trap, i.e., $\eta = 0$. Fig.\,\ref{Fig1} shows that the initial ring-shaped BEC
with $s = 2$ circulation (see (a)) expands toward the center and forms a DQV (see (b)). After that, the core size of the DQV oscillates 
and never decays into two unit quantized vortex, even though the interference pattern manifests in the density distribution
at relatively long time and the BEC experiences quadrupole excitation motion. In contrast to the DQV in a static BEC where the 
vortices are unstable and decay rapidly into two singly quantized vortices \cite{Pethick:2002a}, the vortices in a  
ring-shaped BEC can exist stably. This indicates that the hole-induced inward and outward motion of the rotating ring-shaped 
BEC can stabilize the DQV in a symmetric trap.  In what follows, we demonstrate how the asymmetric geometry of
the harmonic trap affects the vortices.

\section{Splitting and revival}

The initial state of the ring-shaped condensate is chosen to be identical to the one shown in Fig.\,\ref{Fig1}. 
We calculate the dynamics of the condensate after it is released into an asymmetric trap. Some typical density
distributions are shown in Fig.\,\ref{Fig2}, where the value of the isosurface are kept being $1\%$ of the maximum
density of the condensate cloud. In distinction to 
the evolution of the ring-shaped BEC in the rotationally symmetric trap, the DQV here splits into a pair of straight
vortex lines with unit circulation (see Figs.\,\ref{Fig2}(c) and (d)), indicating that the dynamical instability is arisen
from the asymmetric geometry of the trap.  In a later time, the rapidly reducing atoms around the $z$ axis due to the breathing 
motion of the BEC and the resulting local imbalance of the interaction force on the vortices cause two singly quantized vortices 
to recombine into a giant vortex (see Fig.\,\ref{Fig2} (e)), which exhibits a split-and-revival effect. We note that the 
splitting process is very different from those observed in previous work \cite{Mateo:2006a, Huhtamaki:2006a}, where a pair of 
singly charged vortices evolved from a DQV tends to twist around each other in cigar-shaped BECs along the longitudinal direction. 
We find that for an elongated ring-shaped BEC, the twist and interwining of the splitting vortex lines will not happen either, implying
that the collective excitation of the ring-shaped condensate suppresses the higher unstable frequencies and only the lowest one is left, so the splitting takes
place very rapidly along the $z$ axis. The split-and-revival process is also different from the one in Ref.\cite{Rabina:2018a}, where the
giant vortex splits with intertwining but returns nearly to its initial state in an isotropic harmonic trap.   

Our numerical results show that the two vortices are also driven to oscillate due to the collective excitation motion of the BEC. 
For a small $\eta$ (the degree of asymmetry is not sufficiently large), the process of
splitting and revival of the DQV will continue in a relatively long period. The size of the core region of the vortices fluctuates 
with the oscillation of the condensate. After several oscillations, the interference pattern of the matter
wave is manifested clearly. It is shown that the splitting vortices prefer to stay in the interference valley when the interference 
fringes are along the direction of vortex line (see Fig.\, \ref{Fig2} (h)), while the vortex lines are ``cut'' into 
segments when the interference fringes are perpendicular to the vortex lines (see Fig.\, \ref{Fig2} (i)).

\begin{figure}[b]
\centering
\includegraphics[scale = 0.8]{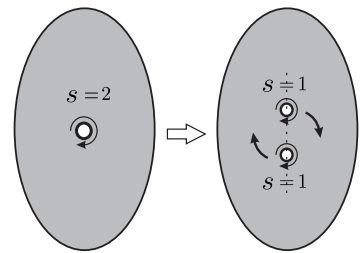} 		        	
\caption{(Color online) Schematic plot of the splitting of 2-charged vortex induced by the asymmetric trap potential.}
\label{Fig3}
\end{figure}

We note that the DQV prefers to split along the elongated direction of the BEC. The conservation of local energy is crucial for such
splitting through dynamical instability. The local energy of $s$-charged vortex is approximately  $\epsilon_v \simeq s^2 \pi
n \frac{\hbar^2}{m} \ln \frac{b}{\xi}$ \cite{Pethick:2002a}, where the spatial variation of local density with the radius $b$ around the vortex is
neglected. $n$ is the background density around the vortex and $\xi$ is the healing length of the BEC. After splitting, $s$ singly
quantized vortices are created and have approximately the energy $\epsilon_{v}^{'} \simeq s \pi n^{'} \frac{\hbar^2}{m} \ln \left(1.464
\frac{b}{\xi} \right)$ where the interaction energy between vortices is neglected \footnote{If the separation $d$ of two singly
quantized vortices is much larger than the healing length $\xi$, this approximation is also valid.}. For $s = 2$ in our case,
two single quantized vortices prefers to stay in the area of $n^{'} = n$ to ensure that the local energy around them is not 
changed greatly in the dynamic process. This results that the 2-charged vortices split into two unit quantized vortices along the elongated direction
of the condensate cloud, where the density varies more slowly than the one in the shorten direction and is closer to the density
in the position of the initial vortices (see Fig.\,\ref{Fig3}).


\begin{figure}[hptb]
\centering
\includegraphics[scale = 0.7]{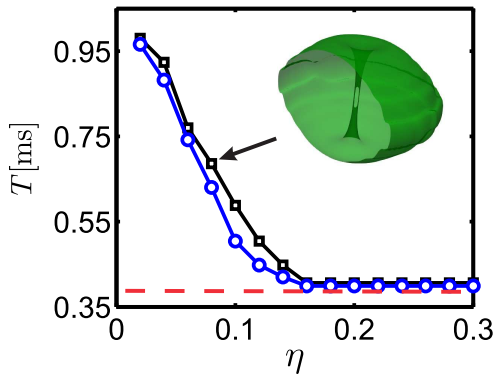} 		        	
\caption{(Color online) The lifetime of a DQV against its decay into two singly quantized vortices, $T$, as a function
of the degree of anisotropy in the trap geometry, $\eta$. The dark squares indicate $g = g_0$ and the blue circles $g = 10 g_0$.
The red dashed line denotes the point of time at which the ring-shaped BEC merges fully into the central region and forms a DQV. 
The insert shows a representative density plot at the critical point of time where the DQV begins to deform into two
singly quantized vortices. We confirm the splitting of a DQV by identifying the structure of two separated vortex
lines in the density isosurface for less than $0.1\%$ of the maximum density.}
\label{Fig4}
\end{figure}

To explore the effect of the asymmetry trap geometry on the splitting of DQVs further, we perform
a series of calculations and show the lifetime of the DQVs, $T$, with respect to the degree of anisotropy, $\eta$,
in Fig.\,\ref{Fig4}. For $\eta < 0.16$, $T$ decays approximately exponentially with increasing $\eta$, while it reaches 
its lower limit at about $\eta = 0.16$. It indicates that when $\eta$ is above the critical value, the DQV is extremely unstable 
and rapidly splits
into  two singly quantized vortices if ever the DQV is formed from the merging ring-shaped BEC. However, for $\eta$ smaller than
the critical value, the lifetime of the DQV in the ring-shaped BEC is dependent of the trap geometry and becomes stable for a
sufficiently small degree of anisotropy. By comparing the blue circles and the dark squares in Fig.\,\ref{Fig4}, one can see that 
for a given $\eta$ smaller than the critical value, a strong interatomic interaction
tends to shorten the lifetime of the DQV, despite it speeds up the merging of ring-shaped BEC and the resulting 
formation of the DQV in an early time.
Since the critical value of $\eta$ is not changed manifestly for different interatomic interaction, this implies that the
geometry of the harmonic trap is a prominent factor for the dynamical instability of the DQV produced in the ring-shaped BEC.
Moreover, We emphasize that the theoretical lifetime here should be much shorter than the realistic experimental observation
when the identical parameters are taken into account. We identify the separation of two vortex lines with its inner structure
of the density isosurface for less than $0.1\%$ of the maximum density but the outer structure near the surface of the
condensate appears still as one core (see Fig.\,\ref{Fig4} insert). By contrast, experiments confirm the DQV splitting through
the visible spatial separation of two vortex lines, e.g., similar to Fig.\ref{Fig2} (f).

\section{Vortex-induced excitation}

\begin{figure}[hptb]
\centering
\includegraphics[scale = 0.09]{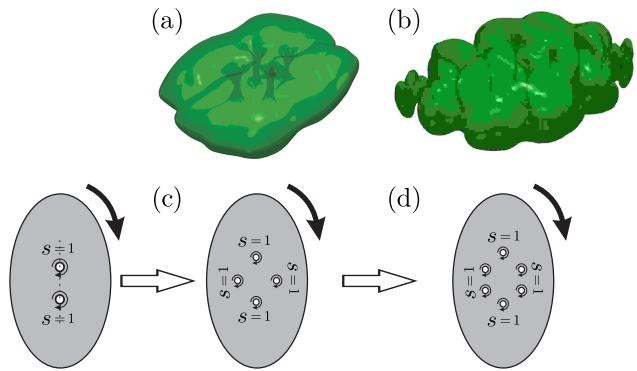} 		        	
\caption{(Color online) The density isosurface of the ring-shaped condensate for $0.1\%$ of the maximum density, corresponding to
$\eta = 0.5$, $g=20g_0$ at $t = 16.2$ms (a) and $\eta = 0.7$, $g=50g_0$ at $t = 11.4$ms (b). (c) and (d) are schematic plots of the
top view of (a) and (b), respectively.}
\label{Fig5}
\end{figure}

   In this part, we further discuss the influence of an increasing interatomic interaction on the dynamic excitation of a ring-shaped
BEC expanding in a fixed anisotropic trap. Due to the $s = 2$ circulation of the whole condensate, the angular momentum of the system
is much larger than the local energy of the DQV. We find that with $g$ increasing up to $20g_0$, the two singly quantized vortices
with $l = 2$ fold rotational symmetry can excite four singly quantized vortices with approximately $4$-fold symmetry
(see Fig.\,\ref{Fig5} (a) and (c)). With stronger interatomic interaction, e.g., $g = 50g_0$, the two singly quantized vortices
excite not only the four vortices but also six quantized vortices with approximately $6$-fold symmetry (see Fig.\,\ref{Fig5} (b)
and (d)). This indicates that the stronger interatomic interaction can produce excitations with higher energy. The dynamic excitations  prefer
to the $\phi$-direction excitations rather than the $r$-direction and $z$-direction. Here we stress that under the same parameters,
the ring-shaped BEC with the $s = 1$ circulation is quite robust and no $l>2$ fold excitations are produced. Note that the 
nonlinear excitation patterns are induced by the multiple vortex and strong interatomic interaction, in distinction to those appear 
in Refs.\cite{Isoshima:2007a, Rabina:2018a, Kuopanportti:2010a} where the low-lying excitation determines the decay pattern of the multiply 
quantized vortex.

\section{Conclusion}
    We have studied the dynamics of the DQV created in ring-shaped Bose-Einstein condensate evolving in
the harmonic traps with varying degree of anisotropic geometry as well as the different interatomic interactions. For an isotropic
trap, the DQV can exist stably, while for the degree of anisotropy below a critical value the life time of the DQV is strongly
dependent on the geometry of the harmonic trap. However, for the degree of anisotropy above the critical value, the DQV becomes
extremely unstable and decay rapidly into two unit vortices once the DQV is formed. Moreover, the finitely large mean-field
interaction has little effect on the critical value for a given ring-shaped BEC. Our simulation suggests that the ring-shaped BEC
can be a desirable candidate to investigate the dynamics of multiply quantized vortex in a controllable way.

\acknowledgments
We thank Wenkai Bai for the inspiring discussions. This work is supported by the NSFC under grants 
Nos. 11775178 and 11775177 and the Major Basic Research Program of Natural Science of Shanxi Province 
under grants Nos. 2017KCT-12 and 2017ZDJC-32.  Y.-J. Lin and D.-W. Wang acknowledges the support 
from NCTS and MoST in Taiwan. Y.-J.L. was supported by a Career Development Award from the Academia Sinica. 

%
%

\bibliography{Refs}

\end{document}